\documentclass[11pt]{article}
\usepackage{graphicx,psfrag,amsmath}
\begin{document}
\newcommand{\beq}{\begin{equation}}
\newcommand{\eeq}{\end{equation}}
\newcommand{\beqa}{\begin{eqnarray}}
\newcommand{\eeqa}{\end{eqnarray}}
\newcommand{\tr}{\mathop{\rm Tr}\nolimits}
\newcommand{\diag}{\mathop{\rm diag}\nolimits}
\newcommand{\mkk}{M_{{\rm KK}} }
\newcommand{\sla}{\slash \hspace{-0.25cm}}
\newcommand{\ratio}{\frac{\mkk}{v}}
\newcommand{\nee}{N_E}
\newcommand{\eff}{A^E_{{\rm eff}}}
\def\lsim{\mathrel{\rlap{\lower4pt\hbox{\hskip1pt$\sim$}}
    \raise1pt\hbox{$<$}}}         %less than or approx. symbol
\def\gsim{\mathrel{\rlap{\lower4pt\hbox{\hskip1pt$\sim$}}
    \raise1pt\hbox{$>$}}}         %greater than or approx. symbol
\begin{titlepage}

\vskip 1.2cm

\begin{center}

{\LARGE\bf Gauge Theories in AdS$_5$ and Fine-Lattice Deconstruction}

\vskip 1.4cm

{\large Andrey Katz and  Yael Shadmi}
\\
\vskip 0.4cm

{\it Physics Department\\
     The Technion---Israel Institute of Technology\\
Haifa 32000, ISRAEL  }\\
     andrey/yshadmi@physics.technion.ac.il\\
\vskip 4pt \vskip 1.5cm

\begin{abstract}
\noindent
The logarithmic energy dependence of gauge couplings in AdS$_5$
emerges almost automatically when the theory is deconstructed
on a coarse lattice. Here we study the theory away from the 
coarse-lattice limit. While we cannot analytically calculate
individual KK masses for a fine lattice, we can calculate the 
product of all non-zero masses. This allows us to write down
the gauge coupling at low energies for any lattice-spacing
and curvature. As expected, the leading log behavior is 
corrected  by power-law contributions, suppressed by the
curvature. We then turn to intermediate energies, and discuss
the gauge coupling and the gauge boson profile in perturbation
theory around the coarse-lattice limit.

\end{abstract}

\end{center}

\vskip 1.0 cm

\end{titlepage}

%%%%%%%%%%%%%%%%%%%%%%%%%%%%%%%%%%%%%%%%%%%%%%%%%%%%%%%%%%%%%%%%%%%%%%%%%%%%%%
\section{Introduction}
%%%%%%%%%%%%%%%%%%%%%%%%%%%%%%%%%%%%%%%%%%%%%%%%%%%%%%%%%%%%%%%%%%%%%%%%%%%%%%
The scale dependence of gauge couplings in AdS$_5$ was studied
using various methods in~\cite{Pomarol:2000hp}-\cite{Goldberger:2003mi}. 
Unlike the situation for flat extra dimensions, 
the running is logarithmic. Thus, even in the 
RS1~\cite{Randall:1999ee} model,
in which the local cut-off on the IR brane is around a TeV, 
it is possible 
to perturbatively extrapolate bulk gauge couplings up to very high scales,
as in four dimensions, and to discuss features of the high-energy theory,
such as unification.
Clearly, the analysis of gauge coupling running in AdS$_5$ is non-trivial.
It involves 
calculating loops in 5-dimensional warped geometry,
or, in a 4-dimensional effective theory description, 
summing over the loop 
contributions of an infinite Kaluza-Klein (KK) tower.
Deconstruction~\cite{Arkani-Hamed:2001ca} produces 
a natural way of regulating these loops. The 5-dimensional gauge theory is 
replaced by a 4-dimensional product-group gauge theory, broken down to the 
requisite gauge group by the VEVs of ``link'' fields. The warped 
geometry 
simply translates to position-, or site-dependent 
VEVs~\cite{Abe:2002rj}-\cite{Falkowski:2002cm}.
For strong warping, 
or equivalently, coarse latticization, the VEVs exhibit a sharp hierarchy,  
so that the problem is even simpler than the in flat case---one can study the 
theory sequentially, starting from the highest VEV and neglecting all smaller 
VEVs at each stage~\cite{Randall:2002qr}. 
In this limit, calculating the masses of KK modes is    
trivial.
Using this approach it was shown in  
references~\cite{Randall:2002qr,Falkowski:2002cm} that the scale dependence 
of the low energy gauge coupling is logarithmic.

However, for energies comparable to the AdS curvature, the coupling
should exhibit the power-law energy dependence typical of flat extra 
dimensions.
 Similarly, power-law effects should also appear in the low-energy
couplings as the curvature is lowered.
In this paper, we explore these effects
for a pure $SU(n)$ gauge theory.

Before going further, it is useful to define more precisely 
the coupling, or rather physical observable, we will study.
An observable associated with some fixed position(s) in the bulk
cannot probe energies all the way up to the Planck scale.
Therefore, as in~\cite{Goldberger:2003mi}, we will imagine having a
quark-antiquark pair localized on the Planck brane.
This quark anti-quark scattering can occur at all energies
up to the Planck scale, and its size gives precisely the
energy-dependent coupling we are after.
In particular, at low-energies, the scattering is mediated
by the zero-mode gauge boson, and so gives the strength
of the coupling of the low-energy gauge group,
which corresponds to the coupling of the unbroken
diagonal gauge-group.

Strictly speaking, in order to obtain the {\it running}
of the gauge coupling, we should calculate it at all energy
scales. 
To do that, we need to compute the KK masses.
But away from the coarse lattice limit, it is
hard to calculate these masses analytically.
We will therefore use perturbation theory
around the coarse lattice limit to calculate
the masses to the first order.
We will then use these results to discuss
the scattering described above, and 
and to define an ``effective gauge boson''
which mediates the scattering.
Thinking in terms of this effective gauge boson sheds
further light on the notion of the position-dependent
regulator brane of~\cite{Randall:2001gc} 
(see also~\cite{Randall:2002qr}).

While we can only calculate the individual KK masses
in perturbation theory, we show that the product of all
nonzero masses can be calculated exactly for finite
curvature and lattice spacing. 
This allows us to obtain the coupling at low energies,
below the lowest KK mass (see section~3).
Indeed, in addition to a logarithmic dependence 
on the UV scale, this coupling contains a term
that scales linearly with the high scale, and is suppressed
by the curvature.

This paper is organized as follows: The set-up is presented 
in Section~2. 
In Section~3 we calculate the coupling below 
the lowest KK mode. 
In Section~4 we discuss the scattering at intermediate
energies.
Appendix A summarizes the calculation of the gauge boson
KK masses and eigenstates. We calculate the product of
non-zero masses in Appendix B.

%%%%%%%%%%%%%%%%%%%%%%%%%%%%%%%%%%%%%%%%%%%%%%%%%%%%%%%%%%%%%%%%%%%%%%%%
\section{Set-Up}
%%%%%%%%%%%%%%%%%%%%%%%%%%%%%%%%%%%%%%%%%%%%%%%%%%%%%%%%%%%%%%%%%%%%%%%%
We consider a pure $SU(n)$ gauge theory in the bulk of the RS1 
model~\cite{Randall:1999ee}. 
For concreteness, we will mostly discuss the non-supersymmetric theory,
but our results carry over trivially to the supersymmetric case.
The deconstructed version of the theory
was discussed in~\cite{Randall:2002qr} and we review it for completeness. 
We write the AdS$_5$  metric as:
\beq \label{metrica}
  ds^2=e^{-2\kappa R\phi}\eta_{\mu \nu}dx^{\mu}dx^{\nu}-R^2d\phi^2~,
\eeq
with Greek indices running over  $0, \ldots , 3$.
The 5d action of the pure $SU(n)$ gauge theory is
\beq\label{5daction}
  S=-\frac14 \int d^5x \sqrt{G} F_{MN}F^{MN}~,
\eeq
where $M,N$ run over $0, \ldots , 4$.
The action~(\ref{5daction}) can be approximated by the deconstructed action 
of an  $SU(n)^N$ 4d gauge theory: 
\beq\label{decaction}
  S=\int d^4x\left[-\frac14 \sum_{a=1}^{N} F_{\mu \nu}^{(a)}F^{\mu \nu (a)}
+\tr \sum_{j=1}^{N-1}\left(D_{\mu}Q_j\right)^\dag \left(D^\mu Q_j\right)
\right]~,
\eeq
where $Q_j$ is a sigma-model scalar field, transforming as $(n, \bar n)$ under 
$SU(n)_j\times SU(n)_{j+1}$.

We assume that the link fields $Q_j$ have vacuum expectation values (VEVs) 
of the form,
\beq \label{VEVmat}
  \langle Q_j \rangle =v_j\diag(1, \ldots, 1)~,
\eeq
breaking the $SU(n)^N$ gauge group down to the diagonal $SU(n)_D$.
In order to describe AdS$_5$ we want these VEVs to scale with the 
warp factor:
 \beq \label{Vevdef}
  v_j=ve^{-j\kappa a}~;\qquad j=1,\ldots (N-1)~. 
\eeq

Classically, the action~(\ref{decaction}) is nothing but a discretization
of the 5d action~(\ref{5daction}) over an $N$-site one-dimensional lattice.
Comparing the two actions, we can relate the 5d lattice spacing, 
$a\equiv R/N$,
and the 5d gauge coupling $g_5$, to the parameters of the 4d theory as follows:
\beq\label{5d4d}
a= {1\over v} \quad,\ \ \ \frac{1}{g_5^2}=\frac{1}{R g_D^2}~.
\eeq
Here $g_D$ is the coupling of the unbroken $SU(n)_D$,
\beq \label{gdiag}
\frac{1}{g_D^2} = \frac{1}{g_1^2} +\cdots + \frac{1}{g_N^2}~.
\eeq
Clearly, at the classical level, we should choose the individual couplings
$g_i$ to have a common value $g_i=g$. 
Since, in AdS, the basic energy scale is position dependent,
and since the coupling $g_j$ is associated with the gauge group at 
the position $j a$,
the scales at which the $g_j$'s attain the common value $g$ should
vary with the warp factor. 
Thus it is natural to define
\beqa\label{coupldef}
g_j( e^{- (j-2) \kappa a} \mkk ) &=& g\ ,\quad \quad  j=2 \ldots N\\
\nonumber g_1( \mkk)&=&g~,
\eeqa
where $\mkk$ is the highest KK mass.

The gauge boson mass matrix is then\footnote{The coupling 
$g$ in front of the mass 
matrix is the gauge coupling of each group, which we can take to be constant 
since we are only working to one-loop}:
\beq \label{massmatrix}
  {\cal M}^2=g^2\left(\begin{array}{ccccc}
           v_1^2&-v_1^2&\ldots&&\\
           -v_1^2&v_1^2+v_2^2&-v_2^2&&\\
           &\ddots&\ddots&\ddots&\\
           &&-v_{N-1}^2&v_{N-1}^2+v_{N}^2&-v_{N}^2\\
           &&\ldots&-v_{N}^2&v_{N}^2\end{array}\right)~.
\eeq
There is one massless mode, given by 
\beq \label{massless}
  A_{ml}=\frac{1}{\sqrt{N}}\left( A_1+A_2+\ldots+A_N\right)~.
\eeq  
It is easy to calculate the remaining masses and mass eigenstates in the flat 
case, $\kappa =0$, as well as for strong curvature. In
the latter case, $\kappa \gg v$, and we can diagonalize the matrix using  
perturbation theory in 
\beq\label{deldef}
  \delta\equiv e^{-2\kappa a}=\frac{v_j^2}{v_{j+1}^2}~.
\eeq  
We do this, up to first order in perturbation theory, in Appendix~A.

The zeroth order result, which was discussed in~\cite{Randall:2002qr}, is
very simple:  $m_j \sim v_j$. In this limit one can study the 
running by turning on the VEVs one at a time. However, we are now interested
in finite curvature for which it is hard to diagonalize the mass matrix
(at least analytically). Still, as we will see in the next section, in order 
to obtain the coupling at low energy, below the lowest KK mass, we only need 
the product of all non-zero masses. As we show in Appendix B, this is given 
by
\beq \label{det}
   m_1^2\ldots m_{N-1}^2=N\prod_{j=1}^{N-1}(gv_j)^2~.
\eeq      
for {\it any} curvature $\kappa$.

%%%%%%%%%%%%%%%%%%%%%%%%%%%%%%%%%%%%%%%%%%%%%%%%%%%%%%%%%%%%%%%%%%%%%%%%%%
\section{The diagonal coupling at low energies}
%%%%%%%%%%%%%%%%%%%%%%%%%%%%%%%%%%%%%%%%%%%%%%%%%%%%%%%%%%%%%%%%%%%%%%%%%%
We will now consider the energy dependence of the gauge coupling,
at energies below the lowest KK mass.
We denote the lowest KK mass by $m_{N-1}$.

At low energies, we have a single $SU(n)$ gauge 
group\footnote{To avoid cumbersome expressions, 
it is convenient to add $n$ fundamentals
(antifundamentals) for $SU(n)_1$ ($SU(n)_N$) so that all 
groups have the same $\beta$-function coefficient.
Then the low energy group has $n$ scalar flavors.}.
Starting at the scale $\mu<m_{N-1}$, we can evolve the coupling
of this diagonal $SU(n)$ up to the high scale $m_1\equiv\mkk$,
\beq \label{gen2}
  \frac{1}{\alpha_{diag}(\mu)}= \frac{1}{\alpha_{diag}(\mkk)} -
\frac{b}{4\pi}\ln \left(\frac{\mkk}{\mu}\right)-
\frac{b}{4\pi}\sum_{j=1}^{N-1}\ln \left(\frac{\mkk}{m_j}\right)~.
\eeq
Here $b=7n/2$ is the one-loop $\beta$-function coefficient
of $SU(n)$ with $n$ flavors, which also equals the contribution
of an adjoint massive vector field.\footnote{In the supersymmetric theory,
$b=2n$.}

Matching the couplings at the highest KK mass, we have
\beq \label{match}
  \frac{1}{\alpha_{diag}(\mkk)}=\sum_{j=1}^{N}\frac{1}{\alpha_j(\mkk)}~.
\eeq
Recall that the couplings $\alpha_j$ are chosen so 
that 
$\alpha_j=\alpha=g^2/(4\pi)$ at the scale
\beq\label{matchscales}
  e^{-(j-2) \kappa a}\, M_{KK}= {M_{KK}\over v_1}\, v_{j-1}= c v_{j-1}~,
\eeq
where we defined the constant $c\equiv M_{KK}/v_1$ for later
convenience. 

Evolving the coupling $g_j$ from this scale up to $\mkk$ we find 
\beq \label{alfarun}
  \frac{1}{\alpha_j(M_{KK})}=\frac{1}{\alpha_j(c v_{j-1} )}
+\frac{b}{4\pi}
\ln \frac{M_{KK}}{c v_{j-1} }~.
\eeq
Combining now eqn.~(\ref{alfarun}) with eqn.~(\ref{gen2})
we find
\beq \label{detrun}
  \frac{1}{\alpha_{diag}(\mu)}=\frac{N}{\alpha}-\frac{b}{4\pi}\ln 
\left (\frac{M_{KK}}{\mu}\right)+\frac{b}{8\pi}\ln 
\left(\frac
{\prod_{j=1}^{N-1}  m_j^2 }
{\prod_{j=1}^{N-1} c^2 v_j^2 }\right)~.
\eeq
Substituting the result~(\ref{det}) in eqn.~(\ref{detrun}) 
and approximating
$N-1$ by $N$ we have
\beq \label{crun}
  \frac{1}{\alpha_{diag}(\mu)}=\frac{N}{\alpha}
-\frac{b}{4\pi}\ln \left(\frac{\mkk/N}{\mu}\right)
-{b\over 8\pi} \left(\ln{\frac{c}{g}}\right)\, N\ 
+\frac{b}{8\pi}\ln{N}~.
\eeq
Note that, in the large $N$ limit, $\mkk$ and $v$ both scale
like $N$, so that $\mkk/v$ and $\mkk/N$ stay finite.
 
The expression~(\ref{crun}) is {\it completely general}: it is valid
for both the warped and flat case.
The curvature only enters this expression through $\mkk$.
In particular, the ratio $c=\mkk/v$, which appears in the
third term of~(\ref{crun}) varies between 2 (for the flat case)
and $\sqrt2$ (in the strongly warped case).

Let us first consider eqn~(\ref{crun}) at non-zero curvature.
Using 
\beq \label{wrpN}
  N=\frac{v}{\kappa}\ln \left( \frac{v}{v_{N-1}}\right)~,
\eeq
we can express the low-energy coupling in terms of the
various energy scales in the problem,
\beqa \label{crunk}
  \frac{1}{\alpha_{diag}(\mu)} &=& \frac{1}{\kappa\alpha_5}
\,\ln{{v\over v_{N-1}} }
-\frac{b}{4\pi}\ln \left(\frac{\mkk/N}{\mu}\right)
+{b\over 4\pi} \,\ln{c}\,
\left(\ln{{v\over v_{N-1}}}\right) {v\over \kappa}
+\nonumber \\
  &+& \frac{b}{8\pi}\ln \left(\frac{v}{\kappa}\right)+\frac{b}{8\pi}\ln \ln 
   \left(\frac{v}{v_{N-1}}\right)~.
\eeqa
The first two terms, which correspond to the tree-level and
one-loop results respectively, 
exhibit a logarithmic dependence
on the high scale $v$ (or equivalently, $\mkk$)
and dominate for large curvature or coarse latticization.
However, the third term, which is suppressed by the curvature,
scales linearly with $v$.
We can also write this term 
in terms of the RS radius $R$ as
\beq
{b\over 4\pi} \,(\ln{c})\, R v \ .
\eeq
reproducing the continuum results 
(see e.g.~\cite{Agashe:2002bx,Goldberger:2003mi}).
Thus, in the limit of strong
curvature, (or coarse latticization), the running is purely
logarithmic as given by the first two terms of~(\ref{crunk}).
But at higher energies, with $v$ comparable to the
curvature, we start to see power-law scaling as 
expected.
It is easy to see the origin of this power-law scaling.
Had we chosen to define the couplings at the scales
of the VEVs, instead of the choice of eqn.~(\ref{coupldef}), this term
would vanish. Thus, the linear dependence on the high scale
is a result of the difference between the matching scale,
$M_{KK}$, and $v$. We can therefore rescale the coupling at
the high scale to absorb this effect. However, if we had
a realistic GUT with several heavy threshold, we would be
left with non-universal term, with linear dependence on the
high scale.

It is worth noting that eqn.~(\ref{det}) explains why the coarse
lattice approximation works so well for the low-energy coupling.
The KK masses only enter this coupling through the product of all
masses divided by the product of all VEVs. Naively, in the coarse lattice 
limit, the masses and the VEVs coincide (up to $g$) so that
the ratio is just $g^N$. But as we see from~(\ref{det}), for any
$N$, the ratio of all masses to all VEVs is $N g^N$. So the
naive coarse lattice result only ``misses'' by $\ln{N}$, which is
not large for the coarse lattice.
Moreover, if one diagonalizes the KK mass matrix more carefully
in the coarse lattice limit (see Appendix~A), one finds
\beq\label{indm}
m_j^2= \frac{j+1}{j} v_j^2~, \ \ \ \ \ j=1,\ldots, N-1 \ .
\eeq
Therefore, the coarse lattice result, including numerical
factors, for the product of all non-zero masses, coincides
exactly with the fine lattice result, when written in terms 
of the VEVs and $N$.

For completeness, we can now turn to the flat case.
Here, $N=R v$,
so eqn.~(\ref{crun}) becomes
\beqa \label{crunf}
  \frac{1}{\alpha_{diag}(\mu)}&=&\frac{R}{\alpha_5}
-\frac{b}{4\pi}\ln \left(\frac{\mkk/N}{\mu}\right)
+{b\over 4\pi} \,(\ln{c})\,
Rv + \nonumber \\
 &+&\frac{b}{8\pi}\ln \left(Rv\right)~. 
\eeqa
At tree-level, the coupling scales linearly with the
compactification scale. 
At the loop level, there is a log piece (the second term)
but this is always smaller than the linear energy dependence
of the third ``power-law'' term. 
In this case, $c=M_{KK}/v=2$, and again, this power law piece
comes from the fact that we matched couplings at the scale
$\mkk$.

%%%%%%%%%%%%%%%%%%%%%%%%%%%%%%%%%%%%%%%%%%%%%%%%%%%%%%%%%%%%%%%%%%%%%%%%%%%%%%
\section{Intermediate energies and the effective gauge boson}
%%%%%%%%%%%%%%%%%%%%%%%%%%%%%%%%%%%%%%%%%%%%%%%%%%%%%%%%%%%%%%%%%%%%%%%%%%%%%%
In the previous section, we only considered the coupling 
of the diagonal $SU(n)$, namely the zero mode.
This is sufficient at low energies, but at intermediate energies,
the heavy KK models become relevant as well,
and so it is not clear which is the relevant coupling.
To see that, it is useful to consider a specific observable
process and study how it varies with the scale.
For example, we can imagine putting a pair of fermions (charged
under the $SU(n)$ group)
on the Planck brane, and study their scattering at various
energies.
At any given energy $E$, the scattering receives contributions from a different
set of gauge KK modes, motivating the introduction
of an ``effective gauge boson'', $\eff$, which mediates the 
scattering at tree level at this energy.\footnote{Note
that we can focus on loop corrections to the gauge
boson propagator only. These loops can be used to calculate the $\beta$
function.}
As a first approximation, we will neglect the contribution
of all KK modes with masses higher than $E$. 
The effective gauge boson at $E$ is then some combination of
KK modes with masses up to $E$.
  
In the deconstructed  theory, the Planck-brane fermion  
translates into a 4d fermion charged under $SU(n)_1$.
At tree level, this fermion only couples to the $SU(n)_1$ gauge
boson $A_1$. 
To find the effective gauge boson and its tree-level coupling,
we need to write $A_1$ in terms of mass eigenstates
and throw out those states with energies above $E$: 
\beq \label{AB}
  |A_1 \rangle=\sum_{j=1}^{N}U_{1j}|B_j\rangle~,
\eeq
where $B_j$ are the mass eigenstates and $U$ denotes the 
diagonalizing matrix of the mass matrix (see Appendix~A).
The $B_j$'s are all massive apart from $B_N$.
Discarding those $B_j$'s with masses above $E$, the
fermion-gauge boson interaction becomes
\beq \label{efflg}
  \Delta {\cal L}=g\sum_{j=\nee+1}^{N} \bar \psi U_{1j} \sla B_j \psi~,
\eeq
where $\nee$ is the number of KK modes with masses above $E$.

We can then define the desired single gauge boson mediating the scattering,
\beq \label{effdef}
  \eff \equiv {\cal N}(E) \,\sum_{j=\nee+1}^{N}U_{1j}B_j~,       
\eeq
where ${\cal N}(E)$ is an energy-dependent normalization constant.  
The interaction term~(\ref{efflg}) then becomes 
\beq \label{intlg}
 \Delta {\cal L}=\frac{g}{{\cal N}(E)}\, \bar \psi \sla A^E_{\rm eff} \psi~,
\eeq 
so the effective coupling is
\beq \label{effcoup}
g_{eff} \equiv \frac{g}{{\cal N}(E)}~.
\eeq
To find ${\cal N}(E)$ we should first diagonalize the KK mass
matrix and find its eigenvalues and eigenstates.
Since we cannot do this analytically for arbitrary curvature and lattice
spacing, we use perturbation theory in $\delta=exp(-2\kappa a)$
of eqn~(\ref{deldef}) around the strong curvature solution
(see Appendix~A for details).

It is useful to discuss the region of validity
of our perturbation theory and the coarse lattice result.
For given $\kappa$ and $R$, the requirement 
that $\delta=exp(-2\kappa a)$ be small gives some upper
limit on $N$. On the other hand, $N$ must be large enough to give
a sensible approximation. 
One case of particular interest is the RS1 model, in which
$\kappa R\sim 30$. 
For $N$ between 30 and 60, $\delta$ varies between roughly
0.1 and 0.3, so we can still trust our perturbation theory,
and deconstruction gives a decent approximation of the
low-lying modes. 

Using the results of Appendix~A we can then compute ${\cal N}(E)$ to 
${\cal O}(\delta)$,  
\beq \label{efc}
  g_{eff}(\nee)=\frac{g}{\sqrt{\nee+1}}\left(1+\delta \, 
\frac{\nee}{2(\nee+1)^3}\right) +{\cal O}(\delta^2)~.
\eeq
In the coarse lattice limit, the KK masses are roughly the VEVs $v_j$, and
\beq \label{whatisne}
  \nee = \frac{1}{\kappa a} \ln \left(\frac{v}{E}\right)~.
\eeq
We would like to know how this result changes
as we go to finer lattices. 
Ideally, if we had exact analytic results for the individual
KK modes, we could simply find the energy $E$ at which
$m_{\nee}=E$. Since we only know the masses to order $\delta$,
the relevant question is how the number of KK modes 
heavier than $E$, $\nee$, changes, as $\delta$ increases.
The result is disappointing. The ${\cal O}(\delta)$ correction
to the mass happens to come with a small numerical coefficient.
Thus, in the regime in which we can trust the calculation,
$\nee$ is not affected. 

Explicitly then, to order $\delta$, we obtain the following result for the 
effective coupling at low and intermediate energies,  
(so $\nee \approx \nee+1$):
\beq \label{intalpha}
\frac{1}{\alpha_{eff}}=\frac{1}{\kappa \alpha_5}\ln \frac{v}{E}
\left(1+e^{-\frac{2\kappa}{v}}
\frac{\kappa^2}{v^2}\frac{1}{\ln^2\frac vE}\right)~,
\eeq
where we have used $\delta\equiv exp(-2k/v)$. 
Apart from a small correction, we have recovered the
well-known classical log behavior of the gauge coupling.
 
To go beyond this classical discussion, one should calculate the loop 
diagrams, taking into account all KK modes.
To do this analytically is prohibitively difficult.
We leave the numerical study of this problem for future work.

Finally, let us turn back to the effective gauge
boson and express it in terms of the original deconstruction gauge
fields
\beq \label{CA}
  |\eff \rangle=\sum_{j=\nee+1}^N\sum_{k=1}^N U_{1j}U_{kj}|A_k\rangle \equiv
  \sum_{k=1}^{N}c_k|A_k\rangle~.
\eeq
Using the results of Appendix~A we find:
\beq \label{rowsum}
  c_k=\begin{cases}
 \frac{1}{\nee+1}+\delta \frac{2\nee}{(1+\nee)^3}&\; 
  \text{if $k=1\ldots \nee$}\\
 \frac{1}{\nee+1}-\delta \frac{\nee^2-\nee}{(1+\nee)^3}&\;
  \text{if $k=\nee+1$}\\
  -\delta \frac{\nee}{(1+\nee)^2}&\; \text{if $k=\nee+2$}\\
   0&\; \text{otherwise}.
      \end{cases} 
\eeq 
The resulting gauge boson profile is plotted
(for $\nee=15$ and $\delta=0.17$) in Figure~1.

This explicitly confirms the observations of~\cite{Randall:2002qr} for the 
gauge boson profile.
\begin{figure}
\begin{center}
\includegraphics[width=70mm]{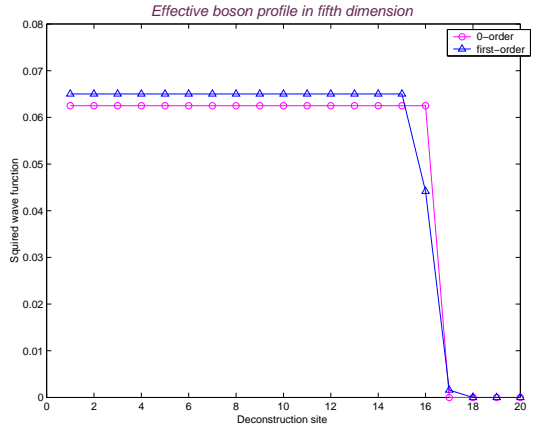}
\caption{The wave-function profile of the effective gauge boson, up to zero 
and first order in perturbation theory in $\delta$. The numerical values are
$\nee=15$ and $\delta=0.17$}
\end{center}
\end{figure} 
The effective gauge boson penetrates deeper
into the fifth dimension as the energy is lowered.
At low energies, the scattering is mediated by the zero mode,
which penetrates all the way down to the TeV brane,
and contains gauge bosons from all sites with equal weights.
At higher energies $E$, in the coarse lattice limit, 
the effective gauge boson is an
equal-weight combination of $\nee$ gauge bosons corresponding to sites
adjacent to the Planck brane.
This picture justifies the energy-dependent cutoff brane
of~\cite{Randall:2001gb}.
Working with a finer lattice, we see that this step-function
profile is actually ``smeared'', and to ${\cal O}(\delta)$, 
the effective gauge boson now receives a small contribution from 
one additional site, namely $A_{\nee+2}$.
At higher orders in perturbation theory, we will
see more sites giving such small contributions,
so that the step function profile becomes smooth.

%%%%%%%%%%%%%%%%%%%%%%%%%%%%%%%%%%%%%%%%%%%%%%%%%%%%%%%%%%%%%%%%%%%%%%%%%%%%%%%
\section{Conclusions and Outlook}
%%%%%%%%%%%%%%%%%%%%%%%%%%%%%%%%%%%%%%%%%%%%%%%%%%%%%%%%%%%%%%%%%%%%%%%%%%%%%%%
In this paper, we studied a deconstructed AdS$_5$ gauge theory,
focusing on  fine latticization, with the inverse lattice
spacing comparable to the curvature. The fact that we obtained 
a general result for the product of all non-zero KK-masses,
allowed us to write down an expression for the coupling
at low energies, which is valid for any curvature and lattice spacing.
Thus, formally at least, this expression interpolates between
the flat case and strongly warped case.

At one-loop, for finite warping, we find, apart from the well-known 
logarithmic piece, a contribution that scales linearly with the
cutoff. This contribution is not universal---it
involves the one-loop beta function coefficient of the theory.

We then discuss the gauge boson propagator at intermediate
energies, 
using perturbation theory around
the coarse lattice limit. 
However, we find that even using this perturbation theory,
it is very hard to analytically calculate loop corrections to the
propagator. 
Furthermore, in the regime in which we can trust the approximation,
the number of KK modes in a given energy interval is not corrected.
Therefore, we cannot see any qualitative modification of
the gauge coupling scale dependence compared to the coarse lattice result. 
It will be interesting to study these issues numerically,
and we leave this analysis for future work.

It will also be interesting to extend our results to
theories containing bulk matter fields, 
and in particular, to consider the deconstruction of
realistic GUT models, with GUT breaking by either boundary conditions 
or by a bulk Higgs field.

\section{Acknowledgements}
We thank A.~Delgado, T.~Gherghetta, and especially Y.~Shirman 
for useful discussions.
We also thank N.~Arkani-Hamed, L.~Randall and N.~Weiner for discussions
that inspired this work.
This research was supported in part by the Israel Science Foundation
(ISF) under grant 29/03, and by the United States-Israel Science Foundation
(BSF) under grant 2002020.

%%%%%%%%%%%%%%%%%%%%%%%%%%%%%%%%%%%%%%%%%%%%%%%%%%%%%%%%%%%%%%%%%%%%%%%%%%%%%%%
\appendix
\section{Mass eigenvalues and eigenstates}
%%%%%%%%%%%%%%%%%%%%%%%%%%%%%%%%%%%%%%%%%%%%%%%%%%%%%%%%%%%%%%%%%%%%%%%%%%%%%%%
Here we diagonalize the gauge boson mass matrix~(\ref{massmatrix}).
We consider a coarse lattice with $\kappa\gsim v$,
and define the hierarchy parameter:
\beq \label{hier}
  \delta \equiv \frac {v_j^2}{v_{j-1}^2}~.
\eeq
In the limit $k\rightarrow \infty$, $\delta$ goes to zero
as does the ratio of adjacent VEVS.
In the following we will obtain the gauge boson masses
and eigenstates as an expansion in $\delta$ around this limit.

We first write ${\cal M}^2$ as the sum of operators:
\beq \label{sumop}
  {\cal M}^2=\sum_{j=1}^{N-1}{\cal M}_j^2~.
\eeq
These operators act  
on the original (site) states as
\beqa\label{opaction}
{\cal M}_j^2 |A_j\rangle &=&v_j^2\left(|A_j \rangle-|A_{j+1}\rangle \right)~; \\{\cal M}_j^2 |A_{j+1}\rangle &=&v_j^2\left(-|A_j \rangle+|A_{j+1}\rangle \right)~;\nonumber \\
 {\cal M}_j^2 |A_l \rangle&=&0 \quad \forall l \ne j,j+1~\nonumber.
\eeqa 
Since we assume a coarse 
lattice, the operator ${\cal M}_1^2$ can be treated as the leading 
operator and all other operators as  perturbations.

The eigenstates of ${\cal M}_1^2$ are: 
\beqa\label{M1eigs}
  \left|B_1\right\rangle&=&\sqrt{\frac12}\left(-\left|A_1\right\rangle
+\left|A_2\right\rangle\right)~;\\
  \left|B_{{m=0}}^1\right\rangle&=&\sqrt{\frac12}\left(\left|A_1\right\rangle
+\left|A_2\right\rangle\right)~,\nonumber
\eeqa 
of masses-squared $2v_1^2$ and zero respectively.
So at this stage, we have $N-1$ degenerate massless
states. This degeneracy is lifted by the operator 
${\cal M}_2^2$.

Iterating this procedure, we find that the operator
${\cal M}_j^2$ , 
treated as a perturbation, removes the degeneracy between the states 
$\left|A_{j+1}\right\rangle$ and $\left|B_{m=0}^{j-1}\right\rangle=
\frac{1}{\sqrt{j}}\sum_{k=1}^{j+1}\left|A_k\right\rangle$. 
The matrix elements of this operator ${\cal M}_j^2$ 
in the sub-space of degeneracy are:
\beqa\label{mjmel}
  \left\langle A_{j+1}\right|{\cal M}_j^2\left|A_{j+1}\right\rangle &=&v_j^2~;\\\left\langle B_{m=0}^{j-1}\right|{\cal M}_j^2\left|B_{m=0}^{j-1}\right\rangle &=&\frac{v_j^2}{j}~;\nonumber\\
  \left\langle A_{j+1}\right|{\cal M}_j^2\left|B_{m=0}^{j-1}\right\rangle &=&-\frac{v_j^2}{\sqrt{j}}~\nonumber.
\eeqa
  
To leading order, the masses are then,
\beq \label{mzerod}
  m_j^2=v_j^2\frac{j+1}{j}~,
\eeq
and solving for the eigenstates we find,
\beqa \label{bjstzer}
  \left|B_j\right\rangle&=&\sqrt{\frac{j}{j+1}}\left(\left|A_{j+1}
\right\rangle-\frac{1}{j}\sum_{k=1}^{j}\left|A_k\right\rangle\right) \qquad j=1\ldots N-1~;\\
  \left|B_{m=0}^{final}\right\rangle&=&\sqrt{\frac1N}\sum_{k=1}^{N}
\left|A_k\right\rangle~.\nonumber
\eeqa
Equations~(\ref{mzerod}) and~(\ref{bjstzer}) give the
infinite-curvature result.
We now want to calculate corrections to these result
for finite curvature. That is, we will now derive the
masses-squared and eigenstates to order ${\cal O}(\delta)$.
The ${\cal O}(\delta)$ correction to each eigenmass can come 
either from  including the second order in ${\cal M}_j^2$, 
or from  first order in 
${\cal M}_{j+1}^2$.
Note that the last eigenstate, $\left|B_{m=0}^{final}\right\rangle$ 
is an exact eigenstate of the mass 
matrix with the zero mass for any curvature, and therefore
should not get corrected.

The second order ${\cal M}_j^2$ correction is
\beq \label{masscorr}
  \Delta m_j^2
=\sum_{k<j}\frac{\left|\left\langle B_j\right|{\cal M}_j^2
\left|B_k\right\rangle \right|^2}{m_j^2-m_k^2}~.
\eeq
with
\beqa
  \left\langle B_j\right|{\cal M}_j^2\left|B_{j-1}\right\rangle 
&=&-v_j^2\frac{\sqrt{(j-1)(j+1)}}{j}~; \\
  \left\langle B_j\right|{\cal M}_j^2\left|B_k\right\rangle 
& = & 0\quad \forall k<j-1~,\nonumber
\eeqa
so there is only one non-zero matrix 
element the numerator of~(\ref{masscorr})numerator sum. 
The denominator contribution is:
\beq
  \frac{1}{m_j^2-m_{j-1}^2}=\frac{-1}{\frac{j}{j-1}v_{j-1}^2}\left(1
+\frac{(j+1)(j-1)}{j^2}\delta+\ldots \right)~,
\eeq
so that eqn.~(\ref{masscorr}) becomes
\beq\label{dm1}
  \Delta m_j^2=\frac{(j-1)(j+1)}{j^2}v_j^4\left(\frac{-1}{\frac{j}{(j-1)}
v_{j-1}^2}+{\cal O}(\delta)\right)=
-\frac{(j-1)^2(j+1)}{j^3}v_j^2\delta+{\cal O}((\delta)^2)~.
\eeq
The contribution of the operator ${\cal M}^2_{j+1}$  is:
\beq\label{dm2}
  \Delta m_j^2=\langle B_j|{\cal M}^2_{j+1}|B_j \rangle=
\frac{j}{j+1}v_j^2\delta~.
\eeq
Together~(\ref{dm1}) and~(\ref{dm2}) give the total first-order 
correction to the mass-squared eigenvalue
\beq
  \Delta m_j^2=v_j^2\left(\frac{j}{j+1}-\frac{(j-1)^2(j+1)}{j^3}\right)\delta~.
\eeq
Adding this to the leading result we find
\beq
  m_j^2=v_j^2\left[\frac{j+1}{j}+\delta \left(\frac{j}{j+1}-
\frac{(j-1)^2(j+1)}{j^3}\right)\right]+{\cal O}(\delta^2)~,
\eeq
or in more convenient form
\beq \label{mr}
   m_j^2=v_j^2\left(\frac{j+1}{j}\right)\left[1+
   \delta \frac{2j^2-1}{j^2(j+1)^2}+{\cal O}(\delta^2)\right]~.
\eeq

The leading order correction to the eigenstates likewise 
comes from two powers of the operator ${\cal M}_j^2$ and 
one power of ${\cal M}_{j+1}^2$,
\beq
  \left|B_j^1\right\rangle =\left|B_j^0\right\rangle+
\sum_{k \ne j}\left|B_k^0\right\rangle \frac{\left\langle B_k^0
\right|{\cal M}_j^2\left|B_j^0\right\rangle}{m_j^2-m_k^2}+
\sum_{k \ne j}\left|B_k^0\right\rangle \frac{\left\langle B_k^0 \right|
{\cal M}_{j+1}^2\left|B_j^0 \right\rangle}{m_j^2-m_k^2}~,
\eeq
where the masses are taken up to zero order in $\delta$.

Using the results of the previous calculation  we obtain the following 
form of mass matrix eigenstates:
\beq\label{basis}
  \left|B_j^1\right\rangle=\left|B_j^0\right\rangle+
\delta \sqrt{\frac{(j-1)^3(j+1)}{j^4}} \left|B_{j-1}^0\right\rangle
-\delta \sqrt{\frac{j^3(j+2)}{(j+1)^4}}\left|B_{j+1}^0\right\rangle 
+{\cal O}((\delta)^2)~.
\eeq
One can easily check that, to ${\cal O}(\delta)$, these eigenstates satisfy
\beq
  \langle B_j|B_k \rangle =\delta_{jk}~.
\eeq

%%%%%%%%%%%%%%%%%%%%%%%%%%%%%%%%%%%%%%%%%%%%%%%%%%%%%%%%%%%%%%%%%%%%%%%%%%
\section{Calculating the product of nonzero masses}
%%%%%%%%%%%%%%%%%%%%%%%%%%%%%%%%%%%%%%%%%%%%%%%%%%%%%%%%%%%%%%%%%%%%%%%%%%%
We now want to calculate the product of all non-zero masses.
To do so, we will use the basis  $|B_j^0\rangle$
of eqn~(\ref{bjstzer}). 
While this basis does not diagonalize the mass matrix
for finite curvature, it is nonetheless useful for
our purposes, since it decouples the
massless mode while leaving a remaining block which is a regular 
matrix with all the required information about the massive KK modes.

To calculate the matrix elements in this remaining block we use the division 
introduced in~(\ref{sumop}). 
In general the action of the operator ${\cal M}_j^2$ on the basis 
$|B_j\rangle$ has the following form:
\beq
  {\cal M}^2_j|B_l\rangle=
\sqrt{\frac{l}{l+1}}\left({\cal M}^2_j|A_{l+1}\rangle-
\frac{1}{l}\sum_{k=1}^{l}{\cal M}_j^2|A_k\rangle \right)~.
\eeq
One can easily see that the second part of this expression 
is non-zero only if $j=l$:
\beq\label{MjAk}
  \sum_{k=1}^{l}{\cal M}_j^2|A_k\rangle=v_j^2\delta_{jl}\left(|A_j\rangle
-|A_{j+1}\rangle \right)~.
\eeq
Using~(\ref{opaction}) and~(\ref{MjAk}) one obtains
\beqa
  {\cal M}_j^2|B_j\rangle&=&v_j^2\sqrt{\frac{j+1}{j}}
\left(-|A_j\rangle+|A_{j+1}\rangle \right)~;\\
  {\cal M}_j^2|B_{j-1}\rangle&=&v_j^2\sqrt{\frac{j-1}{j}}
\left(|A_j\rangle-|A_{j+1}\rangle \right)~;\\
  {\cal M}_j^2|B_l\rangle&=&0\quad\quad \forall l\ne j,j-1~.
\eeqa
Since the operator ${\cal M}_j^2$ is Hermitian,
we conclude that it has only three different non-zero matrix elements in 
the $|B_j\rangle$ basis, which are (for $j\neq1)$,
\beqa
  \langle B_j|{\cal M}_j^2|B_j\rangle&=&\frac{j+1}{j}v_j^2~;\\
  \langle B_{j-1}|{\cal M}_j^2|B_{j-1}\rangle&=&\frac{j-1}{j}v_j^2~;\\
  \langle B_j|{\cal M}_j^2|B_{j-1}\rangle=\langle B_{j-1}|{\cal M}_j^2|B_j
\rangle&=&-\frac{\sqrt{(j-1)(j+1)}}{j}v_j^2~.
\eeqa
For $j=1$ we have only one non-zero element,
\beq
  \langle B_1|{\cal M}_1^2|B_1\rangle=2v_1^2~.
\eeq
Thus, in this basis, the zero mode is explicitly decoupled, 
and the product of KK-masses is just a determinant of the  upper left 
$(N-1)\times (N-1)$ block. 
We can call this $(N-1)\times (N-1)$ matrix the ``reduced matrix'',
or ${\cal M}^2_{red}$, its explicit matrix form is:
\beq
{\cal M}^2_{red}=\left(
 \begin{array}{cccccc}
 2v_1^2+\frac12 v_2^2&-\frac{\sqrt{3}}{2}v_2^2&&&&\\
 -\frac{\sqrt{3}}{2}v_2^2&\frac32 v_2^2+\frac23 v_3^2&\ddots&&&\\
 &\ddots&\ddots&\ddots&&\\
 &&\ddots&\frac{j}{j-1}v_{j-1}^2+\frac{j-1}{j}v_j^2&-
\frac{\sqrt{(j-1)(j+1)}}{j}v_j^2&\\
 &&&-\frac{\sqrt{(j-1)(j+1)}}{j}v_j^2&
\frac{j+1}{j}v_j^2+\frac{j}{j+1}v_{j+1}^2&\ddots\\
 &&&&\ddots&\ddots
 \end{array} \right)
\eeq 
                                               
We then have
\beq \label{dec}
  \det {\cal M}^2_{red}=\det {\cal M}_{N-1}^2 \det \sum_{j=1}^{N-3}{\cal M}_j^2
+{\cal M}^2_{(N-1)(N-1)}\det \sum_{j=1}^{N-2}{\cal M}_j^2~.
\eeq
Using the fact that 
\beq
  \det {\cal M}_j^2=0 \qquad \forall j~,
\eeq
we are left with only one non-zero term on the LHS of~(\ref{dec}). 
Iterating this procedure for the other rows of this matrix 
we finally obtain the desired determinant
\beq
  \det {{\cal M}}_{red}^2=\prod_{j=1}^{N-1}\frac{j+1}{j}v_j^2=
N\prod_{j=1}^{N-1}v_j^2~.
\eeq

\end{document}